\documentclass[aps,prb,preprint,superscriptaddress]{revtex4-1}

\usepackage{times}
\usepackage[T1]{fontenc}
\usepackage[latin1]{inputenc}
\usepackage{epsfig}
\usepackage{graphicx}
\usepackage{graphics}
\usepackage{bm}
\usepackage{hyperref}
\usepackage{amssymb}
\usepackage{amsmath}
\usepackage[english]{babel}
\usepackage{subfig}
\usepackage{color} % cfb
\usepackage{verbatim}

\newcommand{\vvs}{\mathbf{v}_s}
\newcommand{\vvn}{\mathbf{v}_n}
\newcommand{\rr}{\mathbf{r}}

\newcommand{\der}[2]{\frac{\partial {#1}}{\partial {#2}}}

\newcommand{\be}{\begin{equation}}
\newcommand{\ee}{\end{equation}}
\newcommand{\dsp}{\displaystyle}
\newcommand{\beqn}{\begin{eqnarray}}
\newcommand{\eeqn}{\end{eqnarray}}
\newcommand{\refeq}[1]{(\ref{#1})}

\begin{document}

%\preprint{}

%\title{Self--consistent profiles in counterflowing helium II channels}
\title{Large-scale normal fluid circulation in helium superflows}


\author{Luca Galantucci}
\email[]{luca.galantucci@newcastle.ac.uk}
%\thanks{}
\affiliation{Joint Quantum Centre (JQC) Durham--Newcastle, and
School of Mathematics and Statistics, Newcastle University,
Newcastle upon Tyne, NE1 7RU, United Kingdom}
\affiliation{Istituto Nazionale di Alta Matematica, Roma 00185, Italy}

\author{Michele Sciacca}
\email[]{michele.sciacca@unipa.it}
%\thanks{}
\affiliation{Dipartimento di Scienze Agrarie e Forestali, Universit\`a di Palermo}
\affiliation{Istituto Nazionale di Alta Matematica, Roma 00185, Italy}

\author{Carlo F. Barenghi}
\email[]{carlo.barenghi@newcastle.ac.uk}
%\thanks{}
\affiliation{Joint Quantum Centre (JQC) Durham--Newcastle, and
School of Mathematics and Statistics, Newcastle University,
Newcastle upon Tyne, NE1 7RU, United Kingdom}


\date{\today}

\begin{abstract}
We perform fully-coupled numerical simulations of 
helium II pure superflows in a channel, with vortex-line density
typical of experiments. Peculiar to 
our model is the computation of the back-reaction of
the superfluid vortex motion on the normal fluid
and the presence of solid boundaries. 
We recover the uniform vortex-line density experimentally measured 
employing second sound resonators and we show that pure superflow 
in helium~II is associated with a large-scale
circulation of the normal fluid which can be detected using existing
particle-tracking visualization techniques.
\end{abstract}

%\keywords{}

\maketitle

\section{Introduction\label{sec: Intro}}

The problem of velocity profiles in channel flows 
dates back to the pioneering
studies of Poiseuille \cite{poiseuille-1838} and Hagen
\cite{hagen-1839}. Stimulated by genuine curiosity
and industrial purposes, in 1845 Stokes determined that the profile of
an incompressible viscous fluid flowing along a channel
is parabolic\cite{stokes-1845,sutera-skalak-1993} (\textit{Poiseuille profile}). 
Surprisingly, despite half a century of experiments since the first studies performed by Vinen
\cite{vinen-1957a}
%,vinen-1957b,vinen-1957c}
and current important applications
of cryogenics engineering \cite{vansciver-2012},
we still do not know
the profile of superfluid helium (helium~II) flows in a channel.
The difficulty lies in helium II's nature
as the intimate mixture of two fluid components \cite{landau-1941,donnelly-1991}: 
a viscous normal fluid and a inviscid superfluid.
%\cite{tisza-1938,landau-1941,landau-1947}. 
The former can be effectively modelled as an ordinary (classical) 
fluid obeying the
Navier--Stokes equation; the latter is 
akin to textbooks' irrotational inviscid Euler fluid. 
Besides the lack of viscosity,
the key property of the superfluid component is that,
at speed exceeding a small critical value, the potential flow breaks down,
forming a disordered tangle of thin vortex lines of quantized circulation
%$\kappa=10^{-3} \rm cm^2/s$ 
\cite{donnelly-1991,nemirovskii-2013} (unlike classical
fluids whose vorticity is a continuous field). 
These vortex lines couple normal fluid and superfluid via 
a mutual friction force which depends nonlinearly on the
velocity difference between superfluid and normal fluid
and the density of the vortex lines. \cite{barenghi-donnelly-vinen-1983}

The early studies of helium~II channel flows\cite{tough-1982}
lacked the spatial resolution to determine 
flow profiles, and focused instead on global properties 
such as the vortex line density.
The development of innovative low-temperature flow visualization
techniques (based on micron-sized tracers
\cite{zhang-vansciver-2005,bewley-lathrop-sreenivasan-2006} 
or laser-induced fluorescence \cite{guo-etal-2010})
% of metastable helium molecules 
has renewed the interest in flow profiles.
Recent experiments on
thermal counterflow (a regime in which superfluid and normal fluid
move in opposite directions driven by a small heat flux) have shown that
the normal fluid has a tail-flattened laminar profile \cite{marakov-etal-2015}
which undergoes a turbulent
transition \cite{marakov-etal-2015,melotte-barenghi-1998} at larger
heat flux.

\begin{figure}[htbp]
%     \begin{minipage}{0.3\columnwidth}
%      \centering
%       \includegraphics[width=0.95\textwidth]{fig1a.eps}%{./figs/superflow-lam.eps}     
%     \end{minipage}
%    \hspace{0.01\columnwidth}
%     \begin{minipage}{0.3\columnwidth}
%      \centering
%       \includegraphics[width=0.95\textwidth]{fig1b.eps}%{./figs/superflow-turb-fake.eps}       
%     \end{minipage}
%\hspace{0.01\columnwidth}
%     \begin{minipage}{0.3\columnwidth}
%      \centering
%       \includegraphics[width=0.95\textwidth]{fig1c.eps}%{./figs/superflow-turb-right.eps}            %{./figs/vns_0.ps}            %{fig1-bottom.ps}  %./figs/vns_0.ps
%     \end{minipage}
\centering
\includegraphics[width=0.95\textwidth]{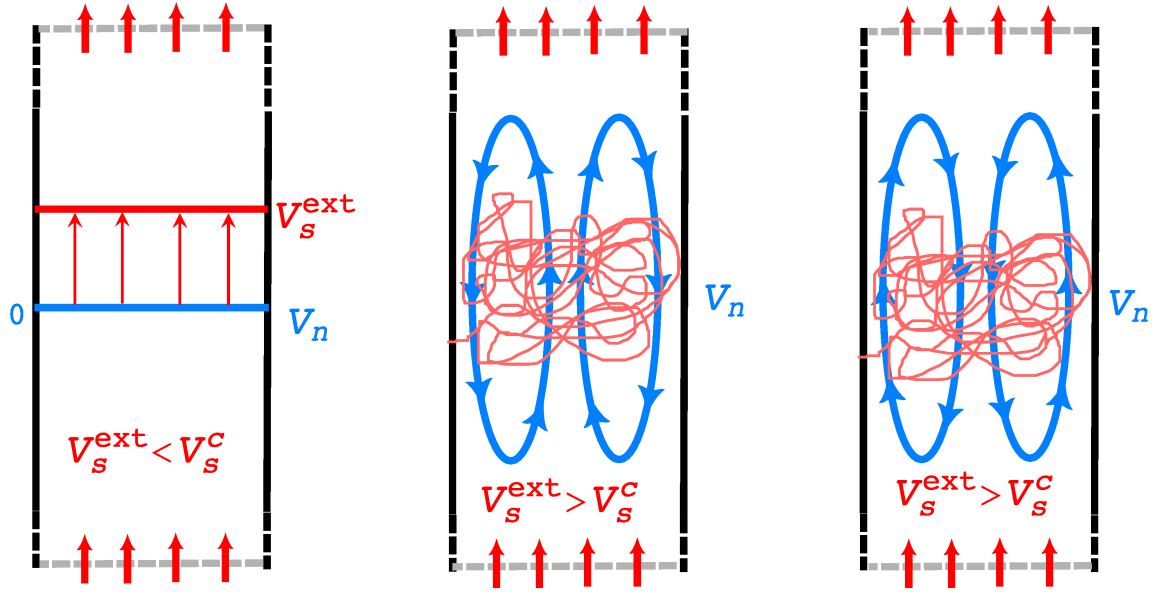}
\caption{(Color online). Schematic illustration of the normal fluid (blue) and
superfluid (red) velocity profiles in a pure superflow channel in a vortex free 
flow (left). (Middle) and (right): conjectures of normal fluid large flow structures 
in presence of a vortex tangle.}
\label{fig:sketch123}
\end{figure}

In this report we focus instead on {\it pure superflow}, another interesting
regime in which the normal fluid is (on the average) at rest
with respect to the channel's walls, while the net superfluid
flow is non-zero. Pure superflow is easily driven thermally or
mechanically by blocking a section of the channel with two superleaks.
\cite{ashton-opatowsky-tough-1981,opatowsky-tough-1981,
baehr-opatowsky-tough-1983,baehr-tough-1984,
babuin-stammeier-varga-rotter-skrbek-2012,babuin-varga-vinen-skrbek-2015}
If the applied superflow $v_s^{ext}$ is less than a small critical velocity
$v_s^c$, the system
is vortex-free, the normal fluid being quiescent and the superfluid
flowing uniformly as schematically shown in Fig.~\ref{fig:sketch123} (left). 
The question which we address \cite{chagovets-skrbek-2008}
is what happens when 
$v_s^{ext} > v_s^c$ and a turbulent tangle of vortices is formed.

Presumably, the vortices via the mutual friction force locally accelerate the 
normal fluid in the direction of $v_s^{ext}$, giving rise to 
large-scale normal fluid circulation whose hypothetical features are shown in
Fig.~\ref{fig:sketch123} (middle) and (right).
%due to the boundary condition of zero net normal flow.
By numerically simulating the two-fluid flow in a dynamically 
self-consistent way 
\cite{galantucci-sciacca-barenghi-2015}
(which allows the normal fluid to affect the vortex lines motion
and viceversa, unlike the traditional approach of Schwarz
\cite{schwarz-1988,baggaley-laizet-2013,baggaley-laurie-2015,
yui-tsubota-2015,khomenko-etal-2015}),
here we show that the normal fluid's circulation pattern coincides with the 
one schematically shown in Fig.~\ref{fig:sketch123} (right). The importance
of this result stems from the belief that the dynamical state of the normal
fluid accounts for the significant observed differences between pure superflow 
and thermal counterflow. 
\cite{tough-ashton-opatowsky-1981,babuin-stammeier-varga-rotter-skrbek-2012,
varga-babuin-skrbek-2015,babuin-varga-vinen-skrbek-2015}
Finally, we show that our prediction can be easily tested using existing 
flow visualization based on solid hydrogen and deuterium tracer particles.
\cite{bewley-lathrop-sreenivasan-2006,lamantia-duda-rotter-skrbek-2013d,
lamantia-skrbek-2014b}

%NOT NECESSARY FOR RAPID COMMUNICATIONS
%The outline of the paper is the following.
%In Section~\ref{sec: model} we briefly describe the 
%two--dimensional model which we use and the main characteristics 
%of the numerical algorithm.
%In Section~\ref{sec: results} we focus on the results and  
%finally Section~\ref{sec: conclusions} summarizes the conclusions.

\section{Model\label{sec: model}}

We consider an infinite two--dimensional channel of width $D$. 
Let $x$ and $y$ be respectively the directions along and across the channel
with walls at $y=\pm D/2$ and periodic boundary conditions imposed at $x=0$ 
and $x=L_x$.  The driving superfluid flow is oriented in the positive $x$ direction.

The superfluid vortices are modelled as $N$ vortex--points of 
circulation $\Gamma_j$ and position $\rr_j(t)=\left ( x_j(t) , y_j(t) \right )$, where $j=1,\cdots N$ and
$t$ is time. Half the vortices have positive circulation $\Gamma_j=\kappa$ and half have negative circulation $\Gamma_j~=~-\kappa$, 
where $\kappa=10^{-3} \rm cm^2/s$ is the quantum of circulation in 
superfluid $^4$He. 

To make connection with experiments we interpret $n=N/(DL_x)$ 
(average number of vortex--points per unit area) as the two--dimensional 
analogue of the three--dimensional vortex--line density $L$,
and relate $L$ to the magnitude of the driving superfluid velocity $v_s^{ext}$
using past experimental data 
\cite{ashton-opatowsky-tough-1981,opatowsky-tough-1981,baehr-opatowsky-tough-1983,baehr-tough-1984,babuin-stammeier-varga-rotter-skrbek-2012}
consistent with Vinen's relation \cite{vinen-1957c} $L^{1/2}=\gamma (v_s^{ext}-v_c)$,
where $\gamma$ is a temperature dependent coefficient and $v_c$ the critical velocity.

%via Vinen's relation\cite{vinen-1957c} $L^{1/2}=\gamma (v_s^{ext}-v_c)$, where
%the coefficient $\gamma (T)$ and the critical velocity $v_c$ have been 
%determined experimentally 
%\cite{ashton-opatowsky-tough-1981,opatowsky-tough-1981,baehr-opatowsky-tough-1983,baehr-tough-1984,babuin-stammeier-varga-rotter-skrbek-2012}.

The vortex points move according to
\cite{schwarz-1988}

\begin{eqnarray}
\displaystyle
\frac{d\rr_j}{dt} & = & \mathbf{v}_s^{0}(\rr_j,t)+\mathbf{v}_{si}(\rr_j,t) \nonumber \\[2mm]
&& + \alpha \, \mathbf{s}_j' \times \left (\vvn(\rr_j,t) - \mathbf{v}_s^{0}(\rr_j,t)-\mathbf{v}_{si}(\rr_j,t) \right ) \nonumber \\[2mm] 
&& + \alpha' \left (\vvn(\rr_j,t) - \mathbf{v}_s^{0}(\rr_j,t)-\mathbf{v}_{si}(\rr_j,t) \right )\label{eq:r_j}
\end{eqnarray} 

\noindent
where $\mathbf{s}_j'$ is the unit vector along vortex $j$ (in the positive 
or negative $z$ direction depending on whether $\Gamma_j$ is positive
or negative), $\alpha$ and $\alpha'$ are temperature
dependent mutual friction coefficients \cite{barenghi-donnelly-vinen-1983},
$\vvn(\rr_j,t)$ is the normal fluid velocity at position $\rr_j$,
$\mathbf{v}_s^{0}(\rr_j,t)$ is the superfluid flow
which enforces the superfluid incompressibility constraint at each channel cross-section and
$\mathbf{v}_{si}(\rr_j,t)$ is the superfluid velocity field induced by 
all the $N$ vortex--points at $\rr_j$:

\begin{equation}
\dsp
\mathbf{v}_{si}(\rr_j,t) = \sum_{k=1 \dots N} \mathbf{v}_{si,k}(\rr_j,t) \, .
\end{equation}

\noindent
To determine the superfluid velocity field induced by the $k$-th vortex 
$\mathbf{v}_{si,k}(\mathbf{x},t)$ we employ a complex--potential--based 
formulation enforcing 
the boundary condition that, at each wall, the superfluid has zero
velocity component in the wall--normal direction \cite{galantucci-sciacca-barenghi-2015}.
%(cfr. Ref.~[\onlinecite{galantucci-sciacca-barenghi-2015}] and Supplementary Material for further details).

The superfluid velocity
$\mathbf{v}_s^{0}(\mathbf{x},t)=\left ( u_{s}^0 (x,t) , 0 \right )$ in 
Eq. \refeq{eq:r_j} is instead obtained by enforcing at each channel cross-section 
the superfluid flow rate determined by the constant driving superfluid velocity 
$\vvs^{ext}=\left ( v_s^{ext} , 0 \right )$, \textit{i.e}

\beqn\label{eq:superflow_condition}
\dsp
%u_s^0(x,t) + \frac{1}{D} \int_{_{-D/2}}^{^{D/2}}\!\!\! u_{si}(x,y,t) dy = u_s^0(x,t) + \langle u_{si} \rangle (x,t) = v_s^{ext}\\
u_s^0(x,t) + \langle u_{si} \rangle (x,t) = v_s^{ext}
\eeqn

\noindent
where $\mathbf{v}_{si}=\left ( u_{si} , v_{si} \right )$ to ease notation and 
$\langle \cdot \rangle$ indicates averaging over channel cross-sections.

To model the creation and the destruction of vortices (mechanisms 
intrinsically three-dimensional) within our two--dimensional model
and to mantain a steady state, we employ the ``numerical vortex reconnection''
procedure described, tested and used in our previous papers 
\cite{galantucci-barenghi-sciacca-etal-2011,galantucci-sciacca-2012,galantucci-sciacca-2014,galantucci-sciacca-barenghi-2015}:
when the distance between two vortex points of opposite circulation becomes 
smaller than a critical value $\epsilon_1$ or when the distance between a vortex point and a 
boundary is less than $\epsilon_2 = 0.5\epsilon_1$, we remove these vortex-points and 
re-insert them into the channel in a random position. 
%In order to assess the dependence of the numerical results on the value of 
%$\epsilon_1$, we have performed numerical simulations varying the value of 
%$\epsilon_1$ by two orders of magnitude: we find that the results are identical 
(refer to Ref.~[\onlinecite{galantucci-sciacca-barenghi-2015}] and Supplementary Material for further insight on this 
numerical reconnection model). 

To investigate the dynamical state of the normal fluid in this two-dimensional model, 
we apply the vorticity-stream function formulation to the incompressible 
Hall-Vinen-Bekarevich-Khalatnikov (HVBK) equations \cite{landau-1941,bekarevich-khalatnikov-1961}
obtaining the following set of equations
\beqn\dsp
\nabla^2 \Psi = -\omega_n \label{eq: poisson psi} \, \, ,
\eeqn
\beqn\dsp 
\der{\omega_n}{t}+\der{\Psi}{y}\der{\omega_n}{x}  & - &  \der{\Psi}{x}\der{\omega_n}{y} \nonumber \\[2mm]
= \nu_n \nabla^2 \omega_n &+& \dsp \frac{1}{\rho_n} \left ( \der{\widetilde{F}^y}{x}  - \der{\widetilde{F}^x}{y} \right ) 
\label{eq: omega_n.3}
\eeqn
where $\widetilde{\mathbf{F}}_{ns}=( \widetilde{F}^x , \widetilde{F}^y )$ is the mutual friction force and 
the stream function $\Psi$ and the normal vorticity $\omega_n$ are defined as follows: 
$\dsp \vvn=\left ( \der{\Psi}{y}, -\der{\Psi}{x} \right )$ , $\dsp \omega_n=\left ( \nabla\times \vvn\right )\cdot \hat{\mathbf{z}}\,\,$
($\hat{\mathbf{z}}$ being the unit vector in the $z$ direction).

%The evolution equation \refeq{eq: omega_n.3} for the normal vorticity $\omega_n'$ is discretized in space
%employing second--order finite differences and its temporal integration 
%is accomplished using the second--order Adams--Bashfort numerical scheme.
%The Poisson equation \refeq{eq: poisson psi} is instead solved in a mixed $(k_x,y)$ space, 
%employing a Fourier--spectral discretization in the periodic 
%$x$--direction and second--order finite differences in the wall--normal direction $y$. 
%The boundary conditions on $\Psi$ and $\omega_n$ are deduced 
%by imposing no--slip boundary conditions on the viscous normal fluid velocity field.

To model the mutual friction force $\mathbf{F}_{ns}$,
we employ the coarse--grained approach of Hall and Vinen \cite{hall-vinen-1956b}
according to which, at lengthscales larger than the average inter-vortex 
spacing $\ell$, the mutual friction assumes 
the following expression:  
\be
\dsp\widetilde{\mathbf{F}}_{ns}=\alpha\rho_s\widehat{\widetilde{\bm{\omega}}}_s \times \left [ \widetilde{\bm{\omega}}_s \times \left ( \widetilde{\mathbf{v}}_n - \widetilde{\mathbf{v}}_s \right ) \right ] + \alpha' \rho_s\widetilde{\bm{\omega}}_s \times \left ( \widetilde{\mathbf{v}}_n - \widetilde{\mathbf{v}}_s \right )\,\, , \label{eq:F_ns}
\ee  
where the symbol $\widetilde{\,}$ over a quantity
indicates that this quantity is coarse--grained. 
At this level of averaging, information about individual vortex lines
is lost, hence it is possible to define continuous macroscopic
superfluid velocity and vorticity fields, $\widetilde{\mathbf{v}}_s$ 
and $\widetilde{\bm{\omega}}_s$ respectively. When computing
coarse-grained quantities, we smooth the vortex distribution using 
a Gaussian kernel to prevent rapid fluctuations of the mutual friction force
\cite{galantucci-sciacca-barenghi-2015} (cfr. Supplementary Material 
for a detailed description of the coarse-graining procedure
and its smearing effects).  

This coarse-grained approach implies to distinguish between the 
\textit{fine} $\left ( \Delta x, \Delta y \right )$ grid on which the normal fluid velocity $\vvn$ is numerically determined, and
the \textit{coarser} $\left (\Delta X,\Delta Y \right )$ grid on which we define the mutual 
friction $\widetilde{\mathbf{F}}_{ns}$. In principle, we would like 
to have $\Delta X$ and $\Delta Y~\gg~\ell $ corresponding to the Hall--Vinen limit; 
in practice, due to computational limitations,
we use $\Delta X \, , \, \Delta Y~>~\ell~>~ \Delta x \, , \, \Delta y$.
%(cfr. Section \ref{sec: results}). 
Once the mutual friction force $\widetilde{\mathbf{F}}_{ns}$ is computed on the coarse grid we interpolate it
on the finer grid via a two--dimensional bi--cubic convolution kernel \cite{keys-1981} 
whose order of accuracy is between linear interpolation and cubic splines orders of accuracy.
It is worth noting that an other method for coupling normal fluid and superfluid motions has also been presented
in past studies \cite{idowu-kivotides-barenghi-samuels-2000,idowu-willis-barenghi-samuels-2000,kivotides-2011},
employing a more fine-scale approach, {\it i. e.} calculating the mutual friction force exerted by each individual vortex
on the normal fluid.

We choose the parameters of the numerical simulations in order to be able to make at least qualitative comparisons with 
the recent experimental superflow studies performed in Prague
\cite{babuin-stammeier-varga-rotter-skrbek-2012,babuin-varga-vinen-skrbek-2015,varga-babuin-skrbek-2015}.  
In particular, we set the width of the numerical channel $D=2.0 \,\, \rm mm$ (comparable to the experimental width
$D_{\rm exp}=7 \div 10 \,\, \rm mm$ \cite{babuin-stammeier-varga-rotter-skrbek-2012,babuin-varga-vinen-skrbek-2015,varga-babuin-skrbek-2015}),
its length $L_x=3D$ and we choose the number of vortices $N$ and the average superfluid driving velocity $v_s^{ext}$
to be consistent with second sound measurements reported in Ref.~[\onlinecite{babuin-stammeier-varga-rotter-skrbek-2012}]:
$N=3072$ and $v_s^{ext}=1.25 \rm cm/s$, leading to $n^{1/2}D=32$. 
It is worth noting that the dimensionless quantity $n^{1/2}D$ is a measure of the superfluid turbulent 
intensity (\textit{i.e.} the larger $n^{1/2}D$, the more intense the superfluid turbulence) 
and it is the relevant quantity to be used when comparing different experiments and when drawing 
parallels between experiments and numerical simulations.
The normal fluid being accelerated by the motion of vortices implies that the 
Reynolds number of the normal fluid flow $\displaystyle Re_n < \frac{v_s^{ext}(D/2)}{\nu_n} = 320$,
far below the critical Reynolds
number for the onset of classical turbulent channel flows $Re_c \approx 5772$.
\cite{orszag-1971}. We reckon therefore that in our
numerical experiment the flow of the normal fluid is still laminar.

\begin{table}
\begin{tabular}{|c|c||c|c|}
\hline
$\,\,\,\,\, D \,\,\,\,\,$ & $2$ & $T$ & $1.7 K$ \\ 
\hline
$L_x$ & $6$ & $\rho_s/\rho_n$ & $3.373$ \\
\hline
$N$ & $3072$ & $\ell$ & $6.25\times 10^{-2}$ \\
\hline
%$\ell$ &  & $\epsilon_1$ & $2.5\times 10^{-3}$ \\
%\hline
\end{tabular}
\caption{Physical and numerical parameters employed in the simulations %and subsequent physical relevant quantities
in dimensionless units}
\label{table: param}
\end{table}

The complete list of parameters employed in our simulation and the physical relevant quantities 
are reported in Table \ref{table: param} and Supplementary Material, 
expressed in terms of the following units of length, velocity and time, respectively:
$\delta_c=D/2=1.0\times 10^{-1}\,\rm cm$, $u_c=\kappa/(2\pi\delta_c)=1.59\times 10^{-3}\,\rm cm/s$, 
$t_c=\delta_c/u_c=62.79\, \rm s$. Hereafter all the quantities which we mention 
are dimensionless, unless otherwise stated. 

For further numerical details concerning the numerical model employed in order to perform the simulations
%(integration methods, timesteps, grids characteristics) 
refer to the Supplementary Material Section.

\section{Results\label{sec: results}}

\begin{figure}[htbp]
\includegraphics[width=0.85\columnwidth]{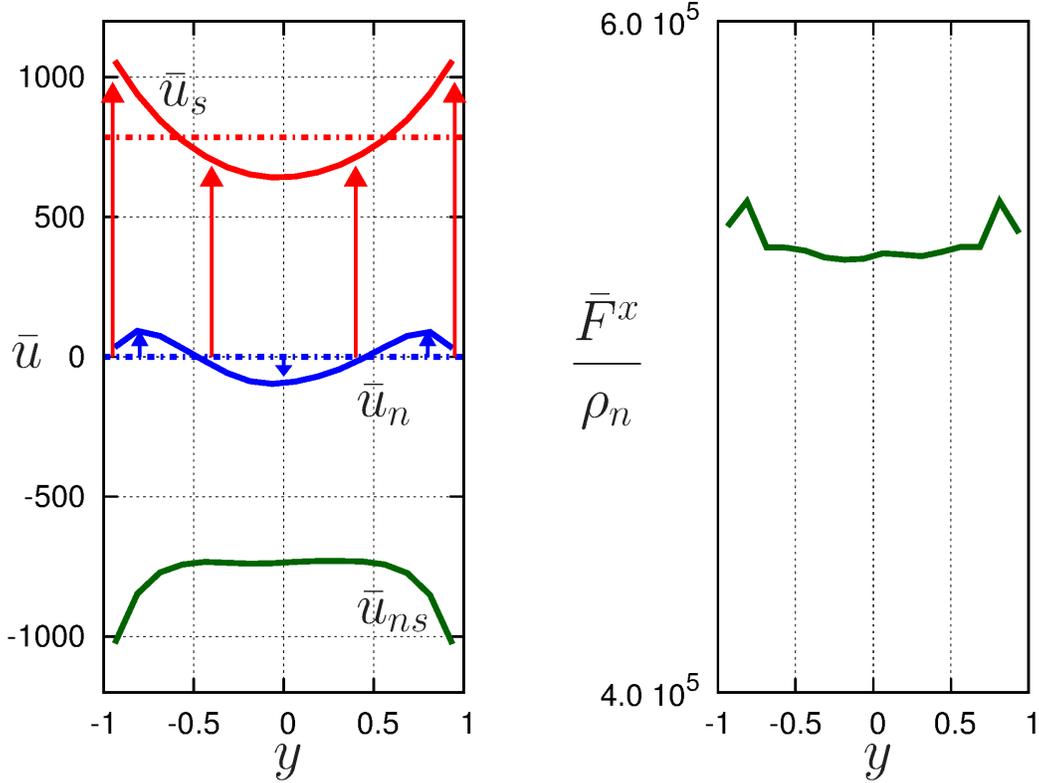}%{./figs/fig_2_small.ps}     
\caption{(Color online). (left): coarse--grained profiles of superfluid velocity
$\overline{u}_s$ (solid red line), normal fluid velocity $\overline{u}_n$ 
(solid blue line) and velocity difference 
$\overline{u}_{ns}=\overline{u}_{n}-\overline{u}_{s}$ (solid green line) at $t=T_{f}$. 
Red and blue dot--dashed lines indicate the initial laminar velocity profiles of
the superfluid and the normal fluid, respectively; 
(right): coarse--grained profile of the streamwise component of the mutual friction force $\overline{F}^x/\rho_n \,$
at $t=T_{f}$. 
\label{fig:velocities}}
\end{figure}

The aim of our numerical simulations is to determine the normal
fluid and superfluid velocity profiles across the channel and the spatial 
distributions of positive and negative vortices in the statistically 
steady--state regime which is achieved after a time interval $T_f$ 
comparable to the viscous eddy turnover time $D^2/\nu_n$. 
To stress that these distributions and profiles
are meant to be coarse--grained over channel 
stripes of size $\Delta Y$,
we use the $\overline{\,\cdot\,}$ symbols. 
The key feature emerging from the numerical simulations is the 
coarse-grained profile of the normal fluid velocity $\dsp\overline{u}_n$
in the steady-state regime, reported in Fig.~\ref{fig:velocities} (left).
The computed profile of $\dsp\overline{u}_n$ shows that the back-reaction of the motion of the superfluid
vortices is effectively capable of driving the motion of the normal fluid whose \textit{local} velocity field
may hence be different from zero. We interpret the computed coarse-grained profile of $\dsp\overline{u}_n$
as the signature of large-scale normal fluid structures 
similar to the ones described schematically in Fig.~\ref{fig:sketch123} (right). It is worth noting that 
in past experimental studies \cite{chagovets-skrbek-2008} the existence of these normal fluid large structures has been
speculated, even though the normal fluid eddies had opposite vorticity (cfr. Fig.~\ref{fig:sketch123} (middle)).
The discriminant element determining the symmetry of the normal fluid flow pattern, 
Fig.~\ref{fig:sketch123} (middle) or Fig.~\ref{fig:sketch123} (right), is the coarse-grained profile of the streamwise
component of the mutual friction force $\overline{F}^x$, illustrated in Fig.~\ref{fig:velocities} (right), 
which is parallel to the driving superfluid velocity $\vvs^{ext}$ and stronger in the near-wall region.
As a consequence, the normal fluid in proximity of the walls is accelerated in the direction of $\vvs^{ext}$, 
while in the central region the normal flow direction is reversed due to the forced re-circulation of the
normal fluid arising from the presence of superleaks and the incompressibilty constraint. It is worth emphasizing
that a similar profile of $\overline{F}^x$ has been recently computed for thermal counterflow 
\cite{galantucci-sciacca-barenghi-2015}.

\begin{figure}[htbp]
\includegraphics[width=0.85\columnwidth]{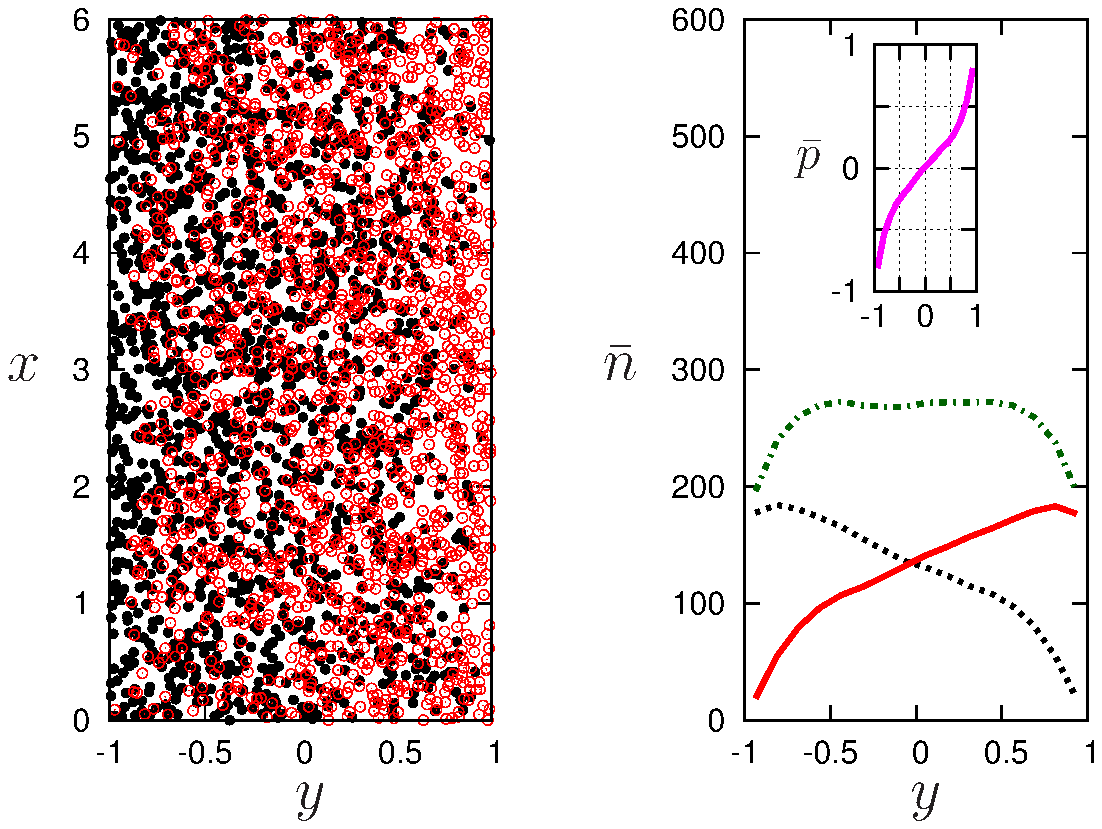}%{./figs/fig_1_small.ps}     
\caption{(Color online). (left): vortex distribution at $t=T_f$, red empty (black filled) 
circles indicate positive (negative) vortices; 
(right): coarse--grained profiles of positive vortex density 
$\overline{n}^+$ (solid red line), negative vortex density 
$\overline{n}^-$ 
(dashed black line) and total vortex density $\overline{n}$ 
(dot--dashed green line) at $t=T_f$. In the inset, we report the corresponding 
coarse--grained profile of the polarization magnitude $\overline{p}$ (solid magenta line). 
\label{fig:vortices}}
\end{figure}

This dynamical equilibrium achieved between the two components of Helium II via the mutual friction coupling is 
characterized by the polarization of the superfluid vortex distribution, which is the other key feature 
arising from the numerical simulations. This spatial configuration %\textit{self-organization}
of the vortex points can be qualitatively observed 
in an instantaneous snapshot of the vortex configuration in the steady state (Fig.~\ref{fig:vortices} (left))
and clearly emerges from the coarse-grained positive and negative vortex density profiles 
illustrated in Fig.~\ref{fig:vortices} (right).
To investigate quantitatively the polarization of the vortex distribution, 
we introduce the coarse--grained polarization vector $\overline{\mathbf p}(y)$ 
defined by \cite{Jou-Sciacca-Mongiovi-2008}
\be 
\displaystyle 
\overline{\mathbf{p}}(y)=\frac{\overline{\bm{\omega}}_s(y)}{\kappa \overline{n}(y)}=\frac{\overline{n}^+(y)-\overline{n}^-(y)}{\overline{n}^+(y)+\overline{n}^-(y)}\hat{\mathbf{z}}\, ,
\ee 
and plot its magnitude $\overline{p}(y)$ in the inset of Fig.~\ref{fig:vortices} (right).
This polarized pattern directly arises from 
the vortex--points equations of motion \refeq{eq:r_j}, where the friction term 
containing $\alpha$ depends on the polarity of the vortex. 
The idealized three-dimensional dynamics corresponding to this polarity-dependent vortex-point motion
is a streamwise flow of expanding vortex-rings lying on planes perpendicular to $\vvs^{ext}$
and drifting in the same direction of the latter. This {\it three-dimensional analogue}
is very similar to the one recently illustrated for thermal counterflow \cite{galantucci-sciacca-barenghi-2015}
and consistent with past three-dimensional analytical \cite{nemirovskii-tsubota-1998}
and numerical \cite{schwarz-1988,baggaley-laurie-2015,yui-tsubota-2015} investigations.
%As it clearly emerges from Fig.~\ref{fig:vortices},
%the circulation of these expanding vortex-rings is oriented in opposite direction with respect to 
%$\vvs^{ext}$.

This polarization of the vortex configuration, which, we stress, is 
\emph{not} complete, \emph{i.e.} $|\overline{p}(y)|<1$,
generates a parabolic coarse--grained 
superfluid velocity profile $\overline{u}_s(y)\sim y^2$ 
which is reported in Fig.~\ref{fig:velocities} (left) and is similar to the
coarse-grained profile computed in recent counterflow simulations \cite{galantucci-sciacca-barenghi-2015}.

It is important to emphasize that our model, although being two--dimensional, recovers
an almost constant profile for the total vortex density $\overline{n}(y)$ across the channel 
(exception made for the near-wall region, cfr. Fig.~\ref{fig:vortices} (right) )
consistent with the recent experimental measurements performed in Prague \cite{varga-babuin-skrbek-2015}.

\section{Conclusions\label{sec: conclusions}}

In conclusion, we have performed self-consistent numerical simulations 
of the coupled normal fluid-superfluid motion in a pure superflow channel,
with values of the vortex-line density typical of recent 
experiments \cite{babuin-stammeier-varga-rotter-skrbek-2012}. 
The main features of our model are that it is dynamically self-consistent
(the normal fluid affects the superfluid and viceversa) unlike the
traditional approach of Schwarz, and that it takes into account the presence
of channel's boundaries. The main approximation of our model is that it is
two-dimensional (as in many studies of classical channel flows).
Nevertheless, we think that this model captures the essential physical
features of the superflow problem: firstly, the model worked well when
applied to counterflow experiments \cite{galantucci-sciacca-barenghi-2015,marakov-etal-2015}; 
secondly, in the superflow problem, it predicts the observed
almost constant vortex density profile
\cite{varga-babuin-skrbek-2015}.

\begin{figure}[htbp]
       %\includegraphics[width=0.15\textwidth]{fig4.eps}%{./figs/superflow-turb-ball.eps}
       %\begin{minipage}{0.3\columnwidth}
      \centering
       \includegraphics[width=0.95\textwidth]{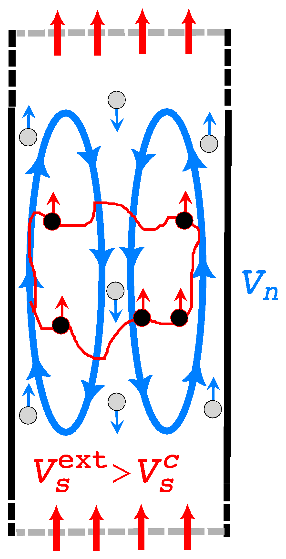}    
     %\end{minipage}
\caption{
(Color online). As in Fig.~\ref{fig:sketch123} (right) but with
tracer particles, which are either free (light-grey circles) or trapped in 
vortices (black circles). 
%It is apparent that near the channel's walls 
%free and trapped particles move in the same direction, whereas near
%the central region free and trapped particles move in opposite
%directions.
}
\label{fig:sketch4}
\end{figure}

Our main prediction is that the normal fluid executes a large-scale 
circulation - see Fig.~\ref{fig:sketch123} (right).
The prediction can be tested 
experimentally employing Particle-Tracking-Velocimetry Visualization 
techniques \cite{bewley-lathrop-sreenivasan-2006,
lamantia-duda-rotter-skrbek-2013d,lamantia-skrbek-2014b}. We know
\cite{poole-barenghi-sergeev-vinen-2005} that at any instant
a tracer particle is either free (in which case it is dragged along by
the normal fluid) or trapped in a vortex line (which,
in first approximation, moves with the applied superflow).
Therefore, in near-wall regions of the channel
we should observe that all particles move in the same direction,
whereas the central region of the channel should contain particles 
moving in both directions. The effect is schematically shown in
Fig.~\ref{fig:sketch4}.

Experimental verification of this effect should
strengthen our general understanding of superfluid hydrodynamics and of
the dynamics of tracer particles in helium~II. It should also open
the way for a better understanding of the onset, steady-state and
decaying state of quantum turbulence.
\cite{tough-ashton-opatowsky-1981,babuin-varga-vinen-skrbek-2015}

\begin{acknowledgments}

LG's work is supported by Fonds National de la Recherche, Luxembourg, 
Grant n.7745104.
LG and MS also acknowledge financial support from the Italian National 
Group of Mathematical Physics (GNFM-INdAM). CFB acknowledges grant
EPSRC EP/I019413/1.
\end{acknowledgments}

%\bibliography{../../mq}
%\break
%\eject

\newpage

%\bibliography{../../../mq}

%\end{document}

%

\end{document}

%\preprint{}

%\title{Self--consistent profiles in counterflowing helium II channels}
%\title{Supplementary Material}

%\maketitle
\begin{center}
\textbf{\underline{{\Large Supplementary Material}}}
\end{center}
\vspace{1cm}

\noindent
In this Supplementary Material Section we describe 
additional, technical characteristics of the numerical method elaborated
for performing the numerical simulations.\\[3mm]

\noindent
\textbf{Superfluid vortices}\\[2mm]
The superfluid velocity $\mathbf{v}_{si,k}(\mathbf{x},t)$ induced by the $k$-th vortex
is determined employing a complex-potential-based formulation enforcing the boundary 
condition that at the channel walls the superfluid velocity has a zero wall-normal component.
The complex potential can be derived using conformal 
mapping \cite{saffman-1992} or, equivalently, using for each vortex an
infinite number of images with respect to the channel walls,
\cite{greengard-1990} leading to the following expression

\beqn\label{eq:F_j}
\dsp
F_k(z,t) = \mp i \frac{h}{2\pi m}\log \frac{\sinh \left [ \frac{\pi}{2D} (z - z_k(t)) \right ]}{\sinh \left [ \frac{\pi}{2D} (z - \overline{z_k}(t) ) \right ] }
\eeqn

\noindent
where $z_k(t)=x_k(t) +iy_k(t)$ is the complex number associated 
to $\rr_k(t)$. The corresponding superfluid velocity
$\mathbf{v}_{si,k}(z,t)=\left ( v_{si,k}^x,v_{si,k}^y \right )$
is obtained from the complex potential in the usual way as 

\begin{equation}
v_{si,k}^x  - iv_{si,k}^y = \frac{dF_k(z,t)}{dz}
\end{equation} 

\noindent
The Lagrangian equation of motion for the superfluid vortices Eq. (1) %\refeq{eq:r_j} 
is integrated in time employing a second-order Adams-Bashfort numerical scheme.\\[3mm]

\noindent
\textbf{Numerical vortex reconnection procedure}\\[2mm]
As illustrated in Section II %\ref{sec: model}
and in Refs.~[\onlinecite{galantucci-barenghi-sciacca-etal-2011,galantucci-sciacca-2012,
galantucci-sciacca-2014,galantucci-sciacca-barenghi-2015}],
we have elaborated a numerical method in order to model 
the superfluid vortex reconnections which are intrinsically 
a three-dimensional phenomenon. The reconnection model 
employed corresponds three-dimensionally to the vortex filament 
method of Schwarz \cite{schwarz-1988} and
correctly describes the fate of two very near
antiparallel vortices (as confirmed by past Gross-Pitaevskii numerical 
studies \cite{koplik-levine-1993}).
In addition, this numerical reconnection procedure prevents from 
the generation of infinitesimal length scales which would 
trigger numerical instabilities and from violating the energy conservation principle.
In order to assess the dependence of the numerical results on the value of 
the distance cut-off $\epsilon_1=2.5 \times 10^{-3}$, we have performed numerical simulations varying the value of 
$\epsilon_1$ by two orders of magnitude: the results obtained are identical.\\[3mm]

\noindent
\textbf{Normal Fluid}\\[2mm]
The evolution equation Eq. (5) %\refeq{eq: omega_n.3} 
for the normal vorticity $\omega_n$ is discretized in space
employing second--order finite differences and its temporal integration 
is accomplished using the second--order Adams--Bashfort numerical scheme.
The Poisson equation Eq. (4) %\refeq{eq: poisson psi} 
is instead solved in a mixed $(k_x,y)$ space, 
employing a Fourier--spectral discretization in the periodic 
$x$--direction and second--order finite differences in the wall--normal direction $y$. 
The boundary conditions on $\Psi$ and $\omega_n$ are deduced 
by imposing no--slip boundary conditions on the viscous normal fluid velocity field.

In the spirit of the coarse--grained description illustrated in Section II, %\ref{sec: model}, 
we define a coarse $\left ( \Delta X, \Delta Y \right )$ 
and a fine $\left ( \Delta x, \Delta y \right )$ grid characterized by number 
of grid--points and spacings listed in Table \ref{table: grid}, satisfying the condition 
$\Delta X, \Delta Y > \ell > \Delta x, \Delta y$. 

\begin{table}[htbp]
\begin{tabular}{||c|c||c|c||}
\hline
\multicolumn{2}{||c||}{Fine grid} & \multicolumn{2}{|c||}{Coarse grid} \\
\hline \hline
$\,\, n_x \,\,$ & $192$ & $\,\, N_x \,\, $ & $48$ \\ 
\hline
$n_y$ & $64$ & $N_y$ & $16$ \\
\hline
$\Delta x$ & $\,\,3.125\times 10^{-2}\,\,$ & $\Delta X$ & $0.125$ \\
\hline
$\Delta y$ & $\,\,3.125\times 10^{-2}\,\,$ & $\Delta Y$ & $\,\, 0.125 \,\, $ \\
\hline
\end{tabular}

\caption{Number of grid--points and spacings in dimensionless units
of the grids employed in the simulations}
\label{table: grid}
\end{table} %\\[3mm]

\noindent
\textbf{Timesteps calculation}\\[2mm]
The coupled calculation of vortex motions and $\vvn$
entails the simultaneous existence of two different timestep stability criteria, one for each motion. Concerning the 
evolution equation Eq. (5) %\refeq{eq: omega_n.3} 
for $\omega_n$, the constraint is set by the normal fluid viscosity \cite{peyret-taylor-1983} 
leading to the restriction $\Delta t_n\le (\Delta x)^2/\nu$. Regarding the motion of the superfluid vortices,
consistently with the numerical reconnection procedure illustrated in Section II, %\ref{sec: model}, 
the integration timestep $\Delta t_v$
for Eq. (1) %\refeq{eq:r_j} 
must satisfy the condition $\Delta t_v\le \epsilon_1/V_{\epsilon_1}$, where $V_{\epsilon_1}$ is the velocity 
of a pair of anti--vortices along their separation vector when separated by a distance equal to $\epsilon_1$. This constraint on $\Delta t_v$ 
prevents from the generation of unphysical small--scale periodic motions (\textit{e.g.} vortex--pairs multiple crossings). 
The value of $\Delta t_v$ employed in our simulation is reported in Table I %\ref{table: param} and 
and we set $\Delta t_n=\Delta t_v=4.0 \times 10^{-6}$. %even though the viscous constraint would have allowed us to set a larger $\Delta t_n$.
Simulations have been performed with smaller timesteps achieving identical results for the statistically steady state. \\[3mm]

\noindent
\textbf{Mutual friction}\\[2mm]

\noindent
We employ the coarse--grained theoretical framework elaborated by Hall and Vinen \cite{hall-vinen-1956b}
according to which, at lengthscales larger than $\ell$, the mutual friction forcing assumes 
the following expression  
\be
\dsp\widetilde{\mathbf{F}}_{ns}=\alpha\rho_s\widehat{\widetilde{\bm{\omega}}}_s \times \left [ \widetilde{\bm{\omega}}_s \times \left ( \widetilde{\mathbf{v}}_n - \widetilde{\mathbf{v}}_s \right ) \right ] + \alpha' \rho_s\widetilde{\bm{\omega}}_s \times \left ( \widetilde{\mathbf{v}}_n - \widetilde{\mathbf{v}}_s \right )\,\, , \nonumber
\ee  
where $\widetilde{\, \cdot \,}$ symbols indicate coarse--grained averaged quantities. 
To prevent rapid fluctuations of the friction at small length--scales, we smooth the vortex distribution using the Gaussian kernel $\Theta_j(\mathbf{r})$
associated to each vortex $j$ according to the following expression 

\beqn\dsp
\Theta_j(\mathbf{r})= \frac{1}{V_j} \dsp e^{\dsp-\frac{|\mathbf{r}-\mathbf{r}_j|^2}{2\ell^2}} \, , 
\eeqn
where $\dsp V_j=\!\!\!\int\limits_{0}^{L_x}\!\int\limits_{-D/2}^{D/2}\!\!\!e^{-\frac{|\mathbf{r}-\mathbf{r}_j|^2}{2\ell^2}} dxdy$. 
Hence, on the basis of Eq. (6), %\refeq{eq:F_ns}, 
the mutual friction force $\widetilde{\mathbf{F}}_{ns}^{p,q}$ averaged on the coarse grid--cell $(p,q)$ is given by the following expression
\beqn\dsp
\widetilde{\mathbf{F}}_{ns}^{p,q} & = & -\alpha \rho_s\, \kappa L^{p,q} \left ( \widetilde{\mathbf{v}}_n^{p,q} - \widetilde{\mathbf{v}}_s^{p,q} \right ) \nonumber \\[5mm]
&& + \alpha' \rho_s \,\Omega^{p,q}\,\hat{\mathbf{z}}\times\left (\widetilde{\mathbf{v}}_n^{p,q} - \widetilde{\mathbf{v}}_s^{p,q}\right )\label{eq:Fns^pq}
\eeqn
where
\beqn\dsp
L^{p,q} &=&\sum_{j=1 \dots N}\,\, \frac{1}{\Delta X \Delta Y} \iint\limits_{(p,q)} \Theta_j(\mathbf{r}) d\mathbf{r} \label{eq: L^pq}\\[5mm]
\Omega^{p,q} &=& \sum_{j=1 \dots N}\,\, \frac{\Gamma_j}{\Delta X \Delta Y} \iint\limits_{(p,q)} \Theta_j(\mathbf{r}) d\mathbf{r} \label{eq: omega^pq}
\eeqn
$\Gamma_j=\pm\kappa$ and the symbol $\dsp\iint\limits_{(p,q)}$ 
denotes the integral over the coarse grid--cell $(p,q)$. 
Physically, $\dsp L^{p,q}$ corresponds to the coarse--grained 
vortex--line density while $\dsp \Omega^{p,q}$ coincides 
with the coarse--grained superfluid vorticity.
Finally, we average $\widetilde{\mathbf{F}}_{ns}^{p,q}$ over the short
time interval $T_{ns}=\Delta X/v_s^{ext}$, the average time 
interval during which a vortex--point moves from a coarse grid--cell
to the neighbouring (cfr. Eq. (1)). %\refeq{eq:r_j}). 

%\bibliography{../../../mq}

%\end{document}

%